\documentclass[prl,twocolumn,aps,superscriptaddress,showpacs,floatfix]{revtex4}
\usepackage{amsmath}
\usepackage{graphicx}

\begin{document}
\title{Beating the Landauer's limit by trading energy with uncertainty}

\author{L. Gammaitoni\cite{lgemail} }
\address{NiPS Laboratory, Dipartimento di Fisica, Universit\'a di Perugia, and Instituto 
Nazionale di Fisica Nucleare, Sezione di Perugia, I-06100 Perugia, Italy}

\date{\today}

\begin{abstract}
According to the International Technology Roadmap for Semiconductors in the next 10-15 years the limits imposed by the physics of switch operation will be the major roadblock for future scaling of the CMOS technology. Among these limits the most fundamental is represented by the so-called Shannon-von Neumann-Landauer limit that sets a lower bound to the minimum heat dissipated per bit erasing operation. Here we show that in a nanoscale switch, operated at finite temperature T, this limit can be beaten by trading the dissipated energy with the uncertainty in the distinguishability of switch logic states. We establish a general relation between the minimum required energy and the maximum error rate in the switch operation and briefly discuss the potential applications in the design of future switches.
\end{abstract}

\maketitle

In the last forty years the semiconductor industry has been driven by its ability to scale down the size of the CMOS-FET\cite{1} switches, the building block of present computing devices, and to increase computing capability density up to a point where the power dissipated in heat during computation has become a serious limitation\cite{2,3}. 
According to the ITRS\cite{4} the limits imposed by the physics of switch operation will be the roadblock for future scaling in the next 10-15 years. The limit on the minimum energy per switching is set at $k_B T \ln(2)$ (approx $10^{-21} J$ at room temperature)\cite{5,6} identified with the so-called Shannon-von Neumann-Landauer\cite{7} (briefly Landauer) limit. 
Power dissipated versus switching speed of devices have been characterized since the seventies\cite{8,9} by a linear scaling rule where micro-fabrication capabilities, through the replacement of bipolar transistors with CMOS, allowed the continuation of the exponential increase trend in information processing capability which has been known as Moore's law\cite{10}. 
However, since 2004 the Nanoelectronics Research Initiative\cite{11}, a US based consortium of Semiconductor Industry Association companies, has launched a grand challenge to address the fundamental limits of the physics of switches. Such limits are mainly represented by the minimum energy and minimum time, required to operate a switch and are estimated by assuming that a two-well, one-barrier model is a valid abstraction for electron transport switching devices. 
In this approach the FET transistor can be thought of as consisting of two wells (source and drain) located at a distance $a$ and separated by a potential energy barrier (channel) of height $E_b$ (see Fig.1). 

\begin{figure}[b]
\includegraphics*[width=8cm]{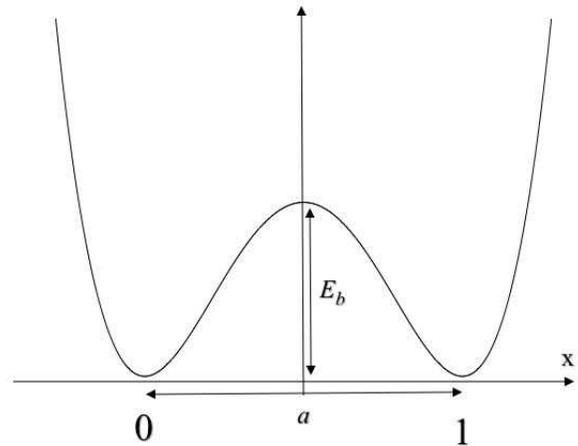}
\caption{Schematic representation of a bistable switch. The switch dynamics can be described in terms of the motion of a particle in a double well potential energy. Each well represents a single logic state (left well, $x<0$, logic state Ò0Ó; right well, $x>0$, logic state Ò1Ó). The two wells are located at a distance $a$ and separated by a potential energy barrier height $E_b$. }
\label{F1} 
\end{figure}

The two logic states Ò0Ó and Ò1Ó are here represented by an electron (or the equivalent information carrier) sitting in the left and right well, respectively. The switching event is obtained by making the electron energetic enough to overcome the potential barrier separating the two states or, what is equivalent, by lowering the potential barrier on the electron side. 
Notwithstanding the simplicity of this model it has been often applied\cite{5} in order to estimate the relevant aspects of the physics of switches. Specifically, the minimum operational energy of the switch is computed by assuming that the barrier height $E_b$ is chosen in order to enable the distinguishability of the two logic states. Such a condition is threaten in fact by unwanted crossings of the potential barrier due to thermally induced (classical) jumps or tunneling (quantum) effects. 
The larger $E_b$ and the distance $a$ between the two wells, the lower the threat to the distinguishability of the two states. Additionally the Heisemberg energy-time indetermination relation is invoked in this context to set a further limit to the barrier height $E_b$. Based on these arguments the authors in \cite{6} have been able to estimate a minimum energy per switching event of the order of few $k_B T \ln(2)$. Clearly such an estimate has a necessarily qualitative character. In order to better highlight the extent of the validity of the arguments used in this estimate we note that in a nanoscale switch in contact with a thermal bath at temperature $T$, the role of fluctuations (thermal or quantum) on the system dynamics can be relevant and it is better accounted by introducing a statistical description of the electron position in terms of a probability density function $p(x)$. In this condition the switch assumes the logic state Ò0Ó with probability $p_0$ measured by the area under $p(x)$ when $x<0$ and assumes the logic state Ò1Ó with probability $p_1$ measured by the area under $p(x)$ when $x>0$. When the potential is symmetric the equilibrium probability distribution dictates: $p_0 = p_1 =1/2$ (Fig 2a).

\begin{figure}[b]
\includegraphics*[width=9cm]{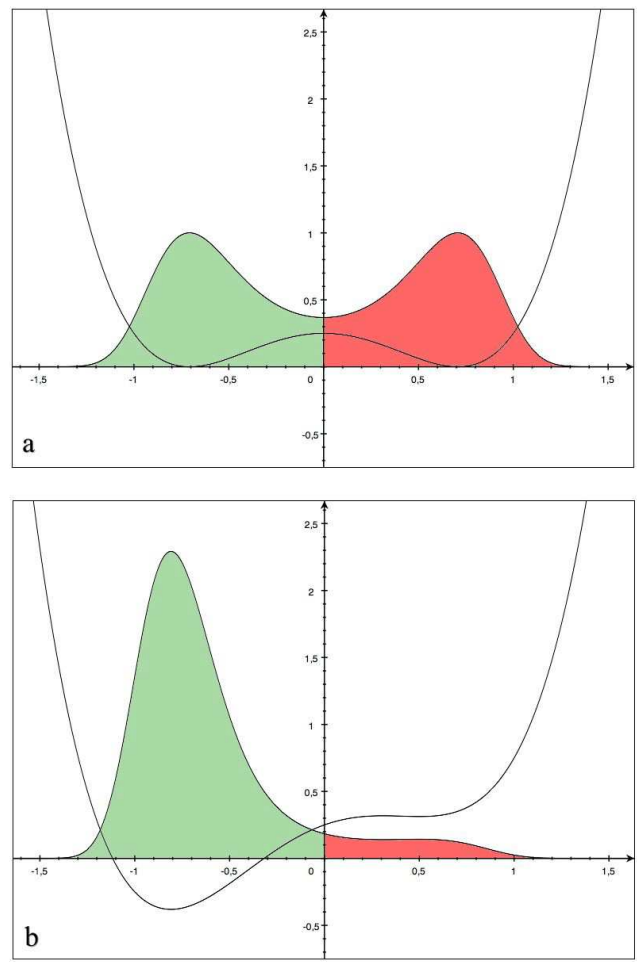}
\caption{Potential energy and probability density function. \b{a}: the probability density function $p(x)$ is plotted together with the potential energy for the symmetric case. The probability of realization of a given logic state is represented by the shaded area below $p(x)$, for Ò0Ó (left well) and for Ò1Ó (right well). \b{b}: Same quantities as in \b{a}, after the reset operation obtained by changing the potential energy with the addition of an external DC signal. The area under the right side can be interpreted as the error probability $P_e$.}
\label{F2} 
\end{figure}

There are two relevant assumptions in the minimum energy estimate above that require a deeper discussion: 

i) the calculation in \cite{5,6} implicitly assumes that the system is far from equilibrium. In fact, in a symmetric system like the one represented in Fig. 1, subjected to thermal and/or quantum fluctuations, regardless the value of $E_b$ and the initial state, if one wait long enough the distiguishability will be lost due to the establishment of equilibrium probability distribution. Thus in order to preserve the distinguishability, the switch has to be operated far from equilibrium. Moreover, during the relaxation at equilibrium process, the switch stochastic dynamics becomes relevant and some attention has to be paid to the statistical features of the fluctuations\cite{12,13}.

ii) The switching operation assumed in \cite{5,6} is a physically irreversible one. In fact an amount of energy equal or larger than $E_b$, required by the electron in the switching process, is unavoidably lost. In other words the tilting of the potential in Fig. 2b is a dissipative operation. Although this is generally the case in standard CMOS technology, recent results\cite{14,15,16} in the studies of energy dissipation at nanoscale could set the road for low-dissipation switching dynamics where an amount of energy smaller than  $E_b$ might be dissipated. However, even in this most favorable case, the minimum switching energy cannot be set to zero due to the logical reversibility issue of the switching operation\cite{17}. In fact, it is known that the erasure of one bit of information (a logically irreversible operation) produces a decrease of the system entropy equal to $k_B \ln(2)$ and thus, if the erasure is operated at temperature $T$, should dissipate at least $Q_L = k_B T \ln(2)$ of energy\cite{7}, a value commonly addressed as the Landauer's limit. It is now a common understanding that such limit holds down to nanoscale\cite{16,19}.

In order to show how to beat the Landauer's limit, letÕs consider a typical logically irreversible switch operation. This is the so-called {\em reset operation} 
where a switch changes from an unknown state to a well-defined state. LetÕs suppose that we want to set the switch to logic state Ò0Ó, corresponding to the system dynamics located in the left well $(x<0)$. Such a reset operation can be achieved by tilting the potential toward left with the result of lowering the potential barrier that separates the two states.  According to Landauer\cite{7} this operation is associated with an unavoidable minimum energy dissipation due to the decrease in entropy as a consequence of the decrease in the number of configuration states available to our system. In fact, following the second principle of thermodynamics a decrease of entropy $\Delta S$ is associated with a minimum energy dissipation $Q = T \Delta S$. The change in entropy is readily computed as follows: initially the switch is in an unknown state thus the number of states available is $2$ (both Ò0Ó and Ò1Ó are plausible). After the reset, the number of available states decreases from $2$ to $1$, thus the associate change in entropy according to Boltzmann is:

\begin{equation}
\label{dsBotlz}
 \Delta S=S_f - S_i = k_B (\ln(1)-\ln(2)) = -k_B \ln(2)
\end{equation}

This results leads immediately to the Landauer's estimate $Q_L= -k_B T \ln(2)$ (minus sign imply that energy gets dissipated into heat). 

In a real nanoscale switch in contact with a thermal bath at temperature $T$ however, the role of fluctuations requires a probabilistic approach where the switch behavior is best addressed in terms of stochastic nonlinear dynamics\cite{18}. In order to fix our ideas we consider the switch operation in terms of the dynamics of a material particle subjected to a nonlinear (bistable) potential (as in Fig. 1) and a fluctuating force. The time evolution of such a particle is described by a proper Langevin equation\cite{Gardiner}:

\begin{equation}
\label{motion}
m  \ddot{x} = -{{dU(x)}\over{dx}} - \gamma \dot{x} + \sigma \xi(t)
\end{equation}

Where $U(x)$ is the bistable potential and $\xi(t)$ represents the fluctuation whose statistical features are connected with the dissipative properties $\gamma$ by a proper fluctuation-dissipation relation. The time evolution of the probability density $p(x,t)$ is usually described in terms of the associated Fokker-Planck equation\cite{Gardiner}. 
In stationary condition the two states: Ò0Ó and Ò1Ó are realized with probability respectively $p_0$ and $p_1$ given by:

\begin{equation}
\label{prob}
p_0 = \int_{-\infty}^{0} {p(x) dx}, \qquad   p_1 = \int_{0}^{+\infty}{p(x) dx}
\end{equation}

where $p(x)$ is the stationary probability density and $p_0+p_1=1$. 

Before the reset operation the potential is symmetric  and thus $p_0 = p_1 = 0.5$. After the reset, the two probabilities change (Fig.2b) and depending on the degree of tilting (i.e. the intensity of the resetting signal) $p_0$ takes a value in the interval $(0.5,1)$ while $p_1=1-p_0$ takes accordingly a value in the interval $(0,0.5)$. Based on our resetting purpose a non-zero $p_1$ can thus be interpreted as the error probability of the reset operation: the larger $P_e=p_1$  the larger the probability of failure. 

\begin{figure}[b]
\includegraphics*[width=8.5cm]{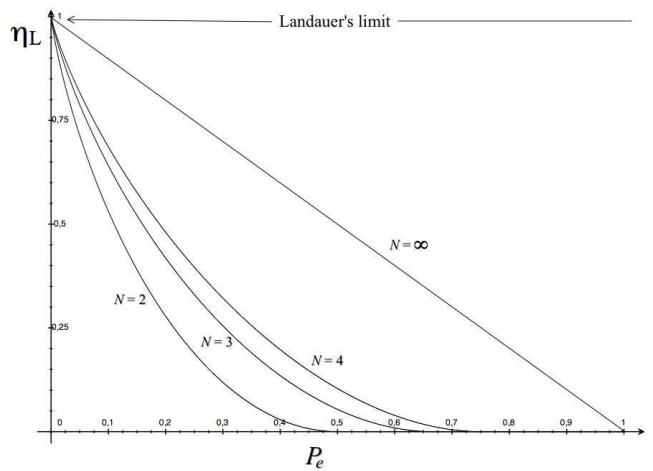}
\caption{Energy ratio $\eta_L$ as a function of the error probability $P_e$. The energy ratio $\eta_L$ represents the fraction of Landauer's energy required to perform the reset operation in the presence of a finite error probability $P_e$. For a bistable switch $(N=2)$ if we accept an error probability of $20\%$ $(Pe=0.2)$ we need to dissipate approx $1/4$ of the minimum required by the Landauer's limit. Multistable switches $(N=3, N=4)$ show an energy ratio $\eta_L$ larger than the bistable one. In the large $N$ limit $\eta_L$ becomes a linear function of $P_e$.}
\label{figure3} 
\end{figure}

Within this description, the change in entropy associated with the reset operation can be computed according to Gibbs as:

\begin{equation}
\label{Gibbs-e}
S = -k_B \sum_i{p_i \ln{p_i}}
\end{equation}
 
where the sum is extended here to the two possible states $i=0,1$.
 The Gibbs entropy before the reset operation $(p_0 = p_1 = 0.5)$ is promptly computed as $S_i=k_B \ln(2)$. After the reset operation, following (\ref{Gibbs-e}) we have:

\begin{equation}
\label{Sf}
S_f(P_e) = -k_B((1- P_e) \ln(1-P_e) + P_e \ln(P_e))
\end{equation}

where we have used $p_0=1-P_e$. Accordingly the energy dissipated during the erasure operation is now a function of the error probability: $Q(P_e) = T \Delta S(P_e)$, i.e.

\begin{eqnarray*}
\label{QPe}
Q(P_e)  & = & -k_B T ((1- P_e) \ln(1-P_e) + P_e \ln(P_e)) +\\
&  & - k_B T \ln(2)
\end{eqnarray*}

In Fig. 3 we plot the energy ratio $\eta_L = Q(P_e)/Q_L$ as a function of the error probability $P_e$. As it is well apparent, for $P_e > 0$ we have $\eta_L <1$, implying that if we are willing to accept a larger-than-zero error probability, we can beat the LandauerÕs limit and perform the resetting operation with an energy toll smaller than $k_B T \ln(2)$. Consistently the zero limit for energy dissipation is reached when $P_e=0.5$, corresponding to the maximum uncertainty (complete undistinguishability according to \cite{5}), i.e. no reset. In the $P_e=0$ limit we regain the Landauer's prediction $Q(0) = Q_L$. 

We point out that the analysis presented here is completely general and does not depend on the switching error mechanism nor on the specific potential landscape typical of different switches. 

A simple generalization of this calculus considers multistable switches where the reset operation takes the switch from an unknown state among $N$ equally probable states to a single, well-defined state. 
In this case the system entropy before the reset operation is $S_i(N) = k_B \ln(N)$ and after the reset operation is

\begin{eqnarray*}
\label{NSf}
S_f(P_e,N) & = & -k_B (1- P_e) \ln(1-P_e) + \\
&  & -k_B P_e (\ln (P_e) - \ln(N-1) )
\end{eqnarray*}

thus the minimum dissipated energy is 
\begin{eqnarray*}
\label{QN}
Q(P_e, N) & = &  - k_B T (1- P_e) ln(1-P_e) + \\
&  &  - k_B T P_e (\ln(P_e)-\ln(N-1)) + \\
&  &  - k_B T \ln(N)
\end{eqnarray*}

The generalized energy ratio $\eta_L = Q(P_e , N)/Q_L(N)$, where now $Q_L(N) = - k_B T \ln(N)$, is also plotted in Fig. 3. As it is well apparent for a given error probability the bistable switch (case with $N=2$) has a lower $\eta_L$ compared to any other multistable switch. In the limit of large $N$  we obtain the asymptotic value 

\begin{equation}
\label{largeN}
\lim_{N\rightarrow \infty} \eta_L = 1- P_e 
\end{equation}

We believe that these results have potential applications in the design of future switches. In fact, computation with switches characterized by $P_e > 0$ is far from being a mere hypothesis. Recently addressed within the paradigm of noise driven switches\cite{13,20,21}, it and has been the topic of a focussed interest in the framework of the so-called stochastic computing\cite{22}, introduced by John Von Newmann\cite{23} since the sixties.

We acknowledge financial support from European Commission (FPVII, G.A. no: 256959, Nanopower and no: 270005, Zeropower) and ONRG Grant N 00014-11-1-0695).

\end{document}